\documentclass{ifacconf}

\usepackage{graphicx}      
\usepackage{natbib}        

\usepackage{amsmath,mathtools}
\usepackage{amsfonts}
\usepackage{xcolor}

\DeclareMathOperator*{\argmin}{arg\,min}

\newcommand{\Vm}{V_\mathrm{m}}

\newcommand{\vx}{\vec{x}}

\def\vec   #1{\mbox{\boldmath $#1$}{}}
\def\scas  #1{\mbox{{\scriptsize{${\rm{#1}}$}}}{}}

\def\ten   #1{\mbox{\boldmath $#1$}{}}

\def\bltr  #1{\mbox{\sffamily{\bfseries{{#1}}}}}

\begin{document}
\begin{frontmatter}

\title{Learning cardiac activation maps from 12-lead ECG with multi-fidelity Bayesian optimization on manifolds\thanksref{footnoteinfo}} 

\thanks[footnoteinfo]{This work was funded by the Leading House for Latin American Region RPG2117 awarded to SP and FSC and the ANID – Millennium Science Initiative Program – NCN19-161 to FSC. This work was also financially supported by the Theo Rossi di Montelera Foundation, the Metis Foundation Sergio Mantegazza, the Fidinam Foundation, the Horten Foundation. We also thank the CSCS-Swiss National Supercomputing Centre for the production grant s1074 awarded to SP.}

\author[CCMC]{Simone Pezzuto}
\author[PEN]{Paris Perdikaris} 
\author[PUC1,PUC2,MIL]{Francisco Sahli Costabal} 

\address[CCMC]{Center for Computational Medicine in Cardiology,
Euler Institute, Università della Svizzera italiana,
Lugano, Switzerland, (e-mail: simone.pezzuto@usi.ch)}
\address[PEN]{Department of Mechanical Engineering and Applied Mechanics, University of Pennsylvania, Philadelphia, Pennsylvania, USA (e-mail: pgp@seas.upenn.edu)}
\address[PUC1]{Department of Mechanical and Metallurgical Engineering, School of Engineering, Pontificia Universidad Cat\'olica de Chile, Santiago, Chile (e-mail: fsc@ing.puc.cl)}
\address[PUC2]{Institute for Biological and Medical Engineering, Schools of Engineering, Medicine and Biological Sciences, Pontificia Universidad Cat\'olica de Chile, Santiago, Chile}
\address[MIL]{Millennium Nucleus for Applied Control and Inverse Problems}

\begin{abstract}
We propose a method for identifying an ectopic activation in the heart non-invasively. Ectopic activity in the heart can trigger deadly arrhythmias. The localization of the ectopic foci or earliest activation sites (EASs) is therefore a critical information for cardiologists in deciding the optimal treatment.
In this work, we formulate the identification problem as a global optimization problem, by minimizing the mismatch between the ECG predicted by a cardiac model, when paced at a given EAS, and the observed ECG during the ectopic activity. 
Our cardiac model amounts at solving an anisotropic eikonal equation for cardiac activation and the forward bidomain model in the torso with the lead field approach for computing the ECG.  We build a Gaussian process surrogate model of the loss function on the heart surface to perform Bayesian optimization. In this procedure, we iteratively evaluate the loss function following the lower confidence bound criterion, which combines exploring the surface with exploitation of the minimum region. We also extend this framework to incorporate multiple levels of fidelity of the model. We show that our procedure converges to the minimum only after $11.7\pm10.4$ iterations (20 independent runs) for the single-fidelity case and $3.5\pm1.7$ iterations for the multi-fidelity case.  We envision that this tool could be applied in real time in a clinical setting to identify potentially dangerous EASs.
\end{abstract}


\end{frontmatter}

\section{Introduction}

\begin{figure*}[tb]
\includegraphics[width=\textwidth]{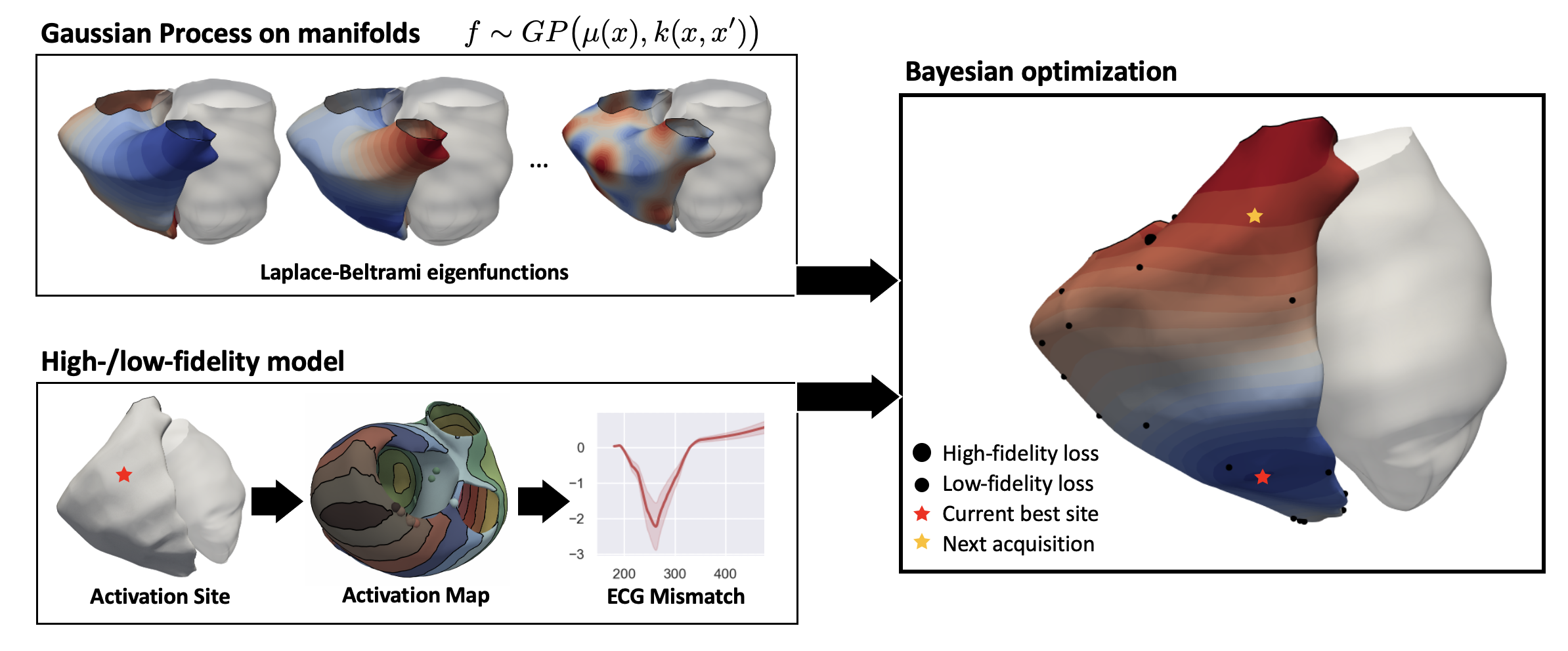}
\caption{Summary of the proposed method to identify an EAS from the ECG. For a tentative
EAS (red star, bottom-left panel), we compute the error in ECG by simulating the activation map and the ECG with an eikonal model. The loss function is obtained from the training set through
GP regression on manifolds, which is based on Laplace-Beltrami eigenfunctions (top-left panel). The loss function is updated using active learning as in Bayesian optimization (right panel).
The best approximation of the best EAS is the global minimum of the loss function.}
\label{fig:intro}
\end{figure*}

Modern cardiology is increasingly shifting towards personalized
approaches inspired by precision medicine, see \cite{peirlinck2021precision}.  The underpinning idea
is to create a so-called digital twin of the patient's heart,
only by using non-invasively acquired data such as those from
cardiac imaging and standard electrocardiography. In this way,
the cardiologist of the future will have the opportunity to
investigate and optimize therapeutic interventions before their
actual implementation on the patient.

The keystone problem in the creation of the digital twin
is the personalization of existing cardiac models from patient-specific
data.  Such models are generally expensive from a computational point of view
and, in order to capture the inter-patient variability, several
parameters need to be optimized, see \cite{grandits2021learning}. A promising approach to reduce the computational burden
is to train a fast surrogate model of the high fidelity model for a selected
set of parameters, e.g., as proposed by~\cite{pagani2021enabling} in the
context of uncertainty quantification. Alternatively, a multi-fidelity
approach is a viable and possibly preferable choice when physiology-based,
low fidelity models are available.  This is actually the case of cardiac
electrophysiology, see e.g.~\cite{Quaglino2019MFMC,sahli2019MF}, since
the eikonal model can effectively approximate
the monodomain model in several cases and at a fraction of the computational cost
of the high fidelity model, as shown by~\cite{neic_efficient_2017,Pezzuto2017Fast}.

The second problem associated with digital twinning is the complexity
of the parameter space.  In this work we focus on the parameters determining
the onset of cardiac activation, the so-called earliest activation sites (EASs).
In healthly conditions, the Purkinje network is the main driver of the ventricular activation, see e.g.~\cite{SahliCostabal2015},
but under pathological conditions like bundle branch block, ventricular tachycardia
and ectopic activation, the EASs are much sparser and more localized, see~\cite{palamara2014computational}.

The problem of localizing the EASs from the 12-lead electrocardiogram (ECG)
has received little attention only until very recently. A common thread of all the recent 
literature is the use of the eikonal model at the base of a fast ECG simulator. 
In this context, the eikonal model is a convenient choice because EASs are
simply Dirichlet boundary conditions to the forward problem, hence easy to
implement. Methodologically, \cite{camps_inference_2021} employ a stochastic method,
namely the Approximate Bayesian Computation (ABC), to determine conduction
velocity and EASs. \cite{gillette_framework_2021} propose instead
a combinatorial approach combined with a parametrization of the heart
to ``flatten'' the parameter space
into a hypercube.  Similarly, \cite{Pezzuto2021ECG} parametrized the
endocardial surfaces in order to apply a derivative-free trust region algorithm to the problem.
Finally, gradient-based methods, respectively based on geodesics
computation and shape derivative, have been proposed by \cite{GranditsGEASI2021}
and \cite{kunisch_inverse_2019}.

In this work, we take a Bayesian perspective to the problem of identifying
a single EAS from the ECG, as for the case of ectopic activity.
Bayesian optimization is a derivative-free
global optimization method that only requires the model evaluation function
to work.  Being based on Gaussian Processes (GPs) for the approximation
of the loss function, it can be adapted to work on manifolds, as proposed
by \cite{lindgren2011}, and
take advantage of multi-fidelity techniques, see e.g.~\cite{kennedy2000predicting}.
Moreover, uncertainty quantification is naturally embedded in the method.


The proposed method, summarized in Figure~\ref{fig:intro},
is explained in details in Section~\ref{sec:methods} and
assessed with numerical experiments on realistic
geometries, reported in Section~\ref{sec:exp}. The outcome of the experiments in discussed in Section~\ref{sec:disc}.

\section{Methods}
\label{sec:methods}


\subsection{Forward ECG model}
\label{sec:forward}

In this work we employ an eikonal-based ECG model. Given an EAS located at $\vec{x}_0\in\mathcal{S}$, where $\mathcal{S}\in\mathbb{R}^3$ is cardiac surface, the eikonal equation provides the activation
map, denoted by $\tau(\vec{x}; \vec{x}_0)$, in the whole myocardium $\Omega\subset\mathbb{R}^3$:
\[
\begin{cases}
\sqrt{\ten{D}(\vec{x})\nabla\tau(\vec{x})\cdot\nabla\tau(\vec{x})} = 1, & \vec{x}\in\Omega, \\
\tau(\vec{x}_0) = 0,
\end{cases}
\]
where $\ten{D}(\vec{x})$ is the conduction velocity tensor. We assume faster conduction in the direction of myocardial fibers, that is $\ten{D}(\vec{x}) = v_t^2 \vec{I} + (v_\ell^2-v_t^2) \vec{l}\otimes\vec{l}$, where $v_\ell$ and $v_t$ are respectively the velocity in the local fiber direction $\vec{l}(\vec{x})$ and in the orthogonal direction.
Next, the transmembrane potential is modeled as traveling wave as follows:
\[
\Vm(\vec{x},t; \vec{x}_0) = U\bigl(t-\tau(\vec{x}; \vec{x}_0)\bigr).
\]
The function $U(\xi)$ represents the action potential, assumed homogeneous
in the domain. Since we are interested in the activation sequence only, a reasonable
approximation smooth step function as follows:
\[
U(\xi) = V_0 + \frac{V_1-V_0}{2}\bigl( \tanh(\xi) + 1\bigr),
\]
where $V_0 = -80\,\mbox{mV}$ and $V_0 = 20\,\mbox{mV}$.

Finally, the ECG is a vector-valued function with
components $V_k(t;\vec{x}_0)$, $k=1,\ldots,12$, that reads as follows:
\[
V_k(t;\vec{x}_0) = \int_\Omega \ten{G}_\mathrm{i}(\vec{x})\nabla\Vm(\vec{x},t;\vec{x}_0)\cdot \nabla Z_k(\vec{x}) \: \mathrm{d}\vec{x},
\]
where $\ten{G}_\mathrm{i}(\vec{x}) = \sigma_{\mathrm{i},t}\ten{I} + (\sigma_{\mathrm{i},\ell}-\sigma_{\mathrm{i},t})\vec{l}\otimes\vec{l}$ is the intra-cellular electric conductivity tensor,
and $Z_k(\vec{x})$ the $k$-th lead field function. See~\cite{Pezzuto2017Fast}
for further details on the implementation.

\subsection{Loss function}

The loss function $\mathcal{F}(\vec{x}_0)$ measures the mismatch between
the simulated ECG $\mathbf{V}(t;\vec{x}_0)$ and the reference ECG $\hat{\mathbf{V}}(t)$
in the least-squares sense. It reads as follows:
\[
\mathcal{F}(\vec{x}_0) = \sum_{k=1}^L \int_0^T \Bigl( V_k(t;\vec{x}_0) - \hat{V}_k(t) \Bigr)^2 \:\mathrm{d}t,
\quad \vec{x}_0 \in\mathcal{S}.
\]
The standard ECG has 12 leads, thus $L=12$.

\subsection{Manifold Gaussian Process}
\label{sec:manifoldGP}

We construct our surrogate model of the objective function using GP regression, see \cite{rasmussen2006gaussian}, a flexible tool for nonlinear Bayesian regression. We assume that we have a data-set of input/output pairs 
$\mathcal{D}= \{(\vec{x}_i, y_i)_{i=1}^{N}\} = \{\vec{X},\vec{y}\}$. 
The input $\vec{X}$ corresponds to $N$ points in the manifold $\mathcal{S}$ that represents the cardiac surface. The output $\vec{y}\in\mathbb{R}^{N}$ contains the corresponding $N$ evaluations of the loss function $\mathcal{F}$ at those locations. 
Our goal is to infer a latent function $f$, such that
\begin{equation}
\vec{y} = f(\vec{X}) + \epsilon,
\end{equation}
where $\epsilon$ is a noise process that may corrupt our observations of $\vec{y}$.
For simplicity, we assume that $\epsilon$ is Gaussian and uncorrelated, 
$\epsilon\sim\mathcal{N}(\vec{0}, \sigma_{n}^{2}\vec{I})$, where $\sigma_{n}^{2}$ is an unknown variance parameter that will be learned from the data.
We assume a zero-mean Gaussian process prior 
\begin{equation}
f(\vec{x}) \sim\mathcal{GP}\bigl(\vec{0},  k(\vec{x},\vec{x}';\theta)\bigr).
\end{equation}

We first identify the optimal set of kernel hyper-parameters and model parameters, $\Theta=\{\theta, \sigma_{n}^{2}\}$, and then use the optimized model to perform predictions at a set of new unobserved locations $\vec{x}^{\ast}$. The covariance kernel function $k(\vec{x},\vec{x}';\theta)$ that depends on a set of hyper-parameters $\theta$ and encodes any prior belief or domain expertise we may have about the underlying function $f$ plays a central role in this process. 
We train our model by
minimizing the negative log-marginal likelihood of the Gaussian process model, see \cite{rasmussen2006gaussian}. Since our likelihood is Gaussian, we can evaluate it analytically in closed form,
\begin{equation} \label{eq:NLML}
\mathcal{L}(\Theta) :=  \frac{1}{2}\log|\vec{K}+\sigma_{n}^{2}\vec{I}| + \frac{1}{2}\vec{y}^{T} (\vec{K}+\sigma_{n}^{2}\vec{I})^{-1}\vec{y} + \frac{N}{2}\log (2\pi),
\end{equation}
where the covariance matrix $\vec{K}\in\mathbb{R}^{N\times N}$ follows from evaluating the kernel function $k(\cdot,\cdot;\theta)$ at the locations of the input training data $\vec{X}$. For the minimization, we adopt a quasi-Newton optimizer L-BFGS with random restarts.
%
Once we have trained our model on the available data, we compute the posterior predictive distribution 
$p(\vec{y}^{\ast}|\vec{x}^{\ast},\mathcal{D})\sim\mathcal{N}(\mu(\vec{x}^{\ast}), \Sigma(\vec{x}^{\ast}))$
at a new location $\vec{x}^{\ast}$ by conditioning on the observed data, 
\begin{equation}
\begin{array}{l@{\hspace*{0.2cm}}c@{\hspace*{0.2cm}}l}
  \mu(\vec{x}^{\ast})  
& =  
& k(\vec{x}^{\ast}, \vec{X}) (\vec{K} + \sigma_{n}^{2}\ten{I})^{-1}\vec{y} \\
  \Sigma(\vec{x}^{\ast}) 
& =  
& k(\vec{x}^{\ast}, \vec{x}^{\ast}) - k(\vec{x}^{\ast}, \vec{X}) (\vec{K} + \sigma_{n}^{2}\ten{I})^{-1} k(\vec{X},\vec{x}^{\ast}) \, ,
\end{array} 
\label{eq:posterior_mean}
\end{equation}
where $\mu(\vec{x}^{\ast})$ and $\Sigma(\vec{x}^{\ast}) $ denote the posterior mean and variance.

A key ingredient of the GP regression is the selection of the kernel function. A common choice is the Mat\'ern kernel, which explicitly allows one to encode smoothness assumptions for the latent functions $f(\vec{x})$, see~\cite{rasmussen2006gaussian}.
In a Euclidean space setting, the kernel function can be explicitly written in terms
of Euclidean distance.

The form based on the Euclidean distance is not suitable to be used on manifolds, as in our case. A na\"ive approach is to replace the Euclidean distance between points
with the geodesic distance on the manifold surface. Even though this approach may work for some cases, there is no guarantee that the resulting covariance will be positive semi-definite, see~\cite{maternGP}, a key requirement for a kernel function. Here, we follow an alternative approach, implicitly
based on the solution of the following stochastic partial differential equation (SPDE), see~\cite{whittle1963stochastic,lindgren2011}:
\begin{equation}\label{eq:spde}
\begin{cases}
(\kappa^2 \mathbf{I} - \Delta)^{\alpha/2} u = \mathcal{W}, & \vx \in \Omega, \\
\mathbf{n}\cdot\nabla (\kappa^2 \mathbf{I} - \Delta)^j u = 0, & \mbox{$\vx \in \partial\Omega$, $j = 0, \ldots, \lfloor\frac{\alpha-1}{2}\rfloor$},
\end{cases}
\end{equation}
where $\alpha=\nu+d/2$ and $-\Delta$ is the Laplace-Beltrami operator on the $d$-dimensional manifold, and $\mathcal{W}$ is the spatial Gaussian white noise on $\Omega$.
When $\Omega = \mathbb{R}^d$, the solution of the fractional SPDE is a Mat\'ern random field evaluated with the Euclidean metric, see~\cite{lindgren2011}.
Therefore, since the SPDE in Eq.~\eqref{eq:spde} trivially generalizes to manifolds with no loss of positive definiteness of the correlation kernel,
thanks to the properties of the pseudo-differential operator, see~\cite{maternGP}.  The correlation function can be explicitly written as follows.
Let $\{ (\lambda_i,\psi_i) \}_{i=0}^\infty$ be the eigenvalue/eigenfunction pairs of the Laplace-Beltrami operator with pure Neumann boundary conditions, that is
\begin{equation}\label{eq:eigen}
\begin{cases}
-\Delta \psi_i = \lambda_i \psi_i & \vx\in\Omega, \\
-\mathbf{n}\cdot\nabla \psi_i = 0, & \vx\in\partial\Omega,
\end{cases}
\end{equation}
for all $i\in\mathbb{N}$. Then, we can represent Matérn-like kernels on manifolds as
\begin{equation}
k(\vx,\vx'; \theta) = \frac{\eta^2}{C}\sum_{\rm i = 0}^{\infty} \left(\frac{1}{\ell^2}+\lambda_i\right)^{-\alpha}\psi_i(\vx)\psi_i(\vx')
\label{eq:materneig}
\end{equation}
where $C$ is a normalizing constant. In practice, the eigendecomposition is truncated to a number $N_\mathrm{eig}$ of pairs.

In this work, we discretize the manifold $\mathcal{S}$ using a triangulated mesh and solve Eq.~\eqref{eq:eigen} using finite element shape functions.
As such, we can obtain the stiffness matrix $\ten{A}$ and mass matrix $\ten{M}$:
\begin{eqnarray}
\ten{A}_{ij} &=& \overset{n_{\scas{el}}}{\underset{\scas{e}=1}{\bltr{A}}} \int_{\mathcal{B}} \nabla N_{ i} \cdot \nabla N_{ j} d\mathcal{B}\\ 
\ten{M}_{ij} &=& \overset{n_{\scas{el}}}{\underset{\scas{e}=1}{\bltr{A}}} \int_{\mathcal{B}}  N_{i} N_{ j} d\mathcal{B},
\end{eqnarray}
where $\bltr{A}$ represents the assembly of the local element matrices, and $N$ are the finite element shape functions. 
Then, we solve the eigenvalue problem:
\begin{equation}
\ten{A}\vec{v} = \lambda \ten{M}\vec{v}
\end{equation}
In practice, to compute the kernel in Eq.~\eqref{eq:materneig} we use a portion of all the resulting eigenpairs, starting from the smallest eigenvalues. We also use the corresponding eigenvectors as the eigenfunctions with $f(\vec{x}_{\rm i}) = \vec{v}_{\rm i}$, where $\rm i$ is the node index at location $\vx_{\rm i}$. Given that the eigenvalue problem is solved only once as a pre-processing step, this methodology provides an efficient way to compute the kernel and the prior in a manifold.

\subsection{Multi-fidelity framework}

In this work, we will assume that we have 2 information sources of different fidelity. We will call the high fidelity, computationally expensive, and hard to acquire information source with the subscript $H$ and the inexpensive, faster to compute, low fidelity source with the subscript $L$. For the problem of interest, the high and low fidelity model are as described in Sec.~\ref{sec:forward}, but with different spatial resolution of the eikonal solver, respectively with grid size of $0.5\,\mbox{mm}$ and $1\,\mbox{mm}$. Now, our data set comes from these two sources $\mathcal{D} = \left\{(\vec{x}_{Li}, y_{Li})^{N_L}_{i=1}, (\vec{x}_{Hi}, y_{Hi})^{N_H}_{i=1} \right\} = \left\{\ten{X}, \vec{y}\right\}$. We will postulate two latent functions $f_H$ and $f_L$, respectively, that are related through an auto-regressive prior, see \cite{kennedy2000predicting}
\begin{equation}
    f_H(\vec{x}) = \rho f_L(\vec{x}) + \delta(\vec{x}).
\end{equation}
Under this model structure, the high fidelity function is expressed as a combination of the low fidelity function scaled by $\rho$, corrected with another latent function $\delta(\vec{x})$ that explains the difference between the different levels of fidelity. Following \cite{kennedy2000predicting}, we assume Gaussian process priors on these latent functions
\begin{eqnarray}
f_L &\sim& \mathcal{GP}\,(\vec{0},k(\vec{x},\vec{x}'; \vec{\theta}_L)), \\
 \delta &\sim& \mathcal{GP}\,(\vec{0},k(\vec{x},\vec{x}'; \vec{\theta}_H)).
\end{eqnarray}
The vectors $\vec{\theta}_L$ and $\vec{\theta}_H$ contain the kernel hyper-parameters of this multi-fidelity Gaussian processes model. The choice of the auto-regressive model leads to a joint prior distribution over the latent functions that can be expressed as
\begin{equation}
\vec{f} =  \left[ 
  \begin{array}{c} \vec{f}_{L} \\ \vec{f}_{H} \end{array} \right] 
\sim \mathcal{N}\left(\left[\begin{array}{c} \vec{0} \\ \vec{0} \end{array} \right],  
\left[ \begin{array}{c c} \ten{K}_{LL} & \ten{K}_{LH}
 \\ \ten{K}_{LH} & \ten{K}_{HH}
  \end{array} 
  \right]\right), 
\end{equation}
with
\begin{equation}
\begin{array}{l@{\hspace*{0.1cm}}c@{\hspace*{0.1cm}}
              l@{\hspace*{0.1cm}}l@{\hspace*{0.1cm}}
              c@{\hspace*{0.1cm}}l@{\hspace*{0.1cm}}l}
  \ten{K}_{LL} 
& = & & k_{L}&(\vec{X}_{L},\vec{X}_{L}';\theta_{L}),  \\
  \ten{K}_{LH} 
& = &\rho & k_{L}&(\vec{X}_{L},\vec{X}_{H}';\theta_{L}), \\
  \ten{K}_{HH} 
& = &\rho^2 & k_{L}&(\vec{X}_{H},\vec{X}_{H}';\theta_{L}) 
& + & k_{H}(\vec{X}_{H},\vec{X}_{H}';\theta_{H})  \,.
\label{eq:MF_K}
\end{array}
\end{equation}

The global covariance matrix $\ten{K}$ of this multi-fidelity Gaussian process model has a block structure corresponding to the different levels of fidelity, where $\ten{K}_{HH}$ and $\ten{K}_{LL}$ model the spatial correlation of the data observed in each fidelity level, and $\ten{K}_{LH}$ models the cross-correlation between the two levels of fidelity. We also have kernel parameters for the different levels of fidelity. We again use the Mat\'ern as described in Sec.~\ref{sec:manifoldGP},  which results in parameters $\vec{\theta}_H = (\eta_H, \ell_H)$, and $\vec{\theta}_L = (\eta_L, \ell_L)$. Again, we find the optimal parameters by minimizing the negative log-marginal likelihood described in \eqref{eq:NLML}.

\subsection{Bayesian Optimization}

To minimize the mismatch between the signals predicted by our model and the observed ones we will use Gaussian process regression to perform Bayesian optimization. In this setting, we will start by computing the objective function at small number of locations ($N = 10$) in the manifold. With this data, we will train a GP regressor to predict the objective function in the entire manifold. Then, we will select the next point in the manifold to compute the objective function as:
\begin{equation}
    \vec{x}^{N+1} = \argmin_{\vec{x} \in \mathcal{S}}\Bigl(\mu(\vec{x}) - 2\sqrt{\Sigma(\vec{x})}\Bigr),
\end{equation}
where $\mu$ and $\Sigma$ are the posterior mean and variance defined in \eqref{eq:posterior_mean}. This is known as the lower confidence bound criterion, see \cite{snoek2012practical}, and combines the exploration of regions of high uncertainty (large variance $\Sigma$) and exploitation of regions that are potentially close to the minimum (low mean $\mu$). We solve this optimization problem by brute force, which involves evaluating the Gaussian process in all the nodes of the mesh that represents the manifold. Once we have acquired the value of the objective function we add it to our dataset $\{\vec{X}, \vec{y}\}$ and repeat this process until $\vec{x}$ matches the true location.

In the multi-fidelity setting, we employ the same approach. The only difference is that we start with $N_H = 5$ high fidelity samples, which are complemented with $N_L = 35$. (The choice is such that the multi-fidelity training cost matches the low fidelity one.) In this approach, we have a better initial approximation of the objective function of the manifold, which should speed up the optimization process. We only acquire high fidelity samples using the lower bound criterion and keep the amount of low fidelity samples fixed.

\section{Numerical Experiments}
\label{sec:exp}

\begin{figure}[tbh]
\includegraphics[width=\columnwidth]{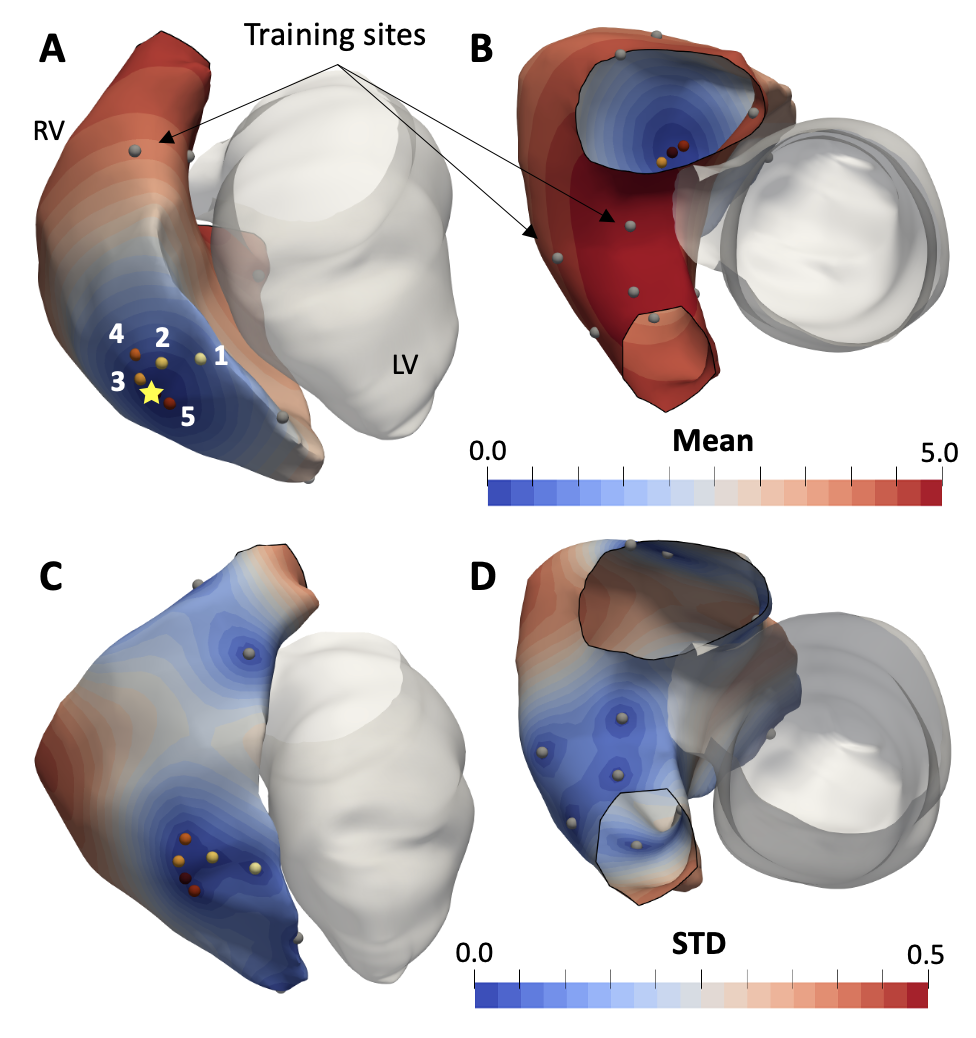}
\caption{Single-fidelity experiment. Top row (A-B panels): mean of the GP, with training sites indicated in gray and numbered iterations. The star indicates the true site.
Bottom row (C-D, views as above): standard deviation of the GP.}
\label{fig:single}
\end{figure}

We present here a comparison of the single-fidelity (SF) and multi-fidelity (MF)
approach.  In both cases, the ground truth $\vec{x}^*$ was located in the mid-free
wall of the right ventricle. The corresponding ECG $\hat{\mathbf{V}}(t)$ was
obtained from the model in Sec.~\ref{sec:forward}, paced at $\vec{x}^*$. Figure~\ref{fig:single} shows an example of the optimization for SF
alone. The total number of iterations was 6, all except the first one very close
to the true minimum.  The Figure also displays the mean and the standard derivation
of the GP.

\begin{figure}[htb]
\includegraphics[width=\columnwidth]{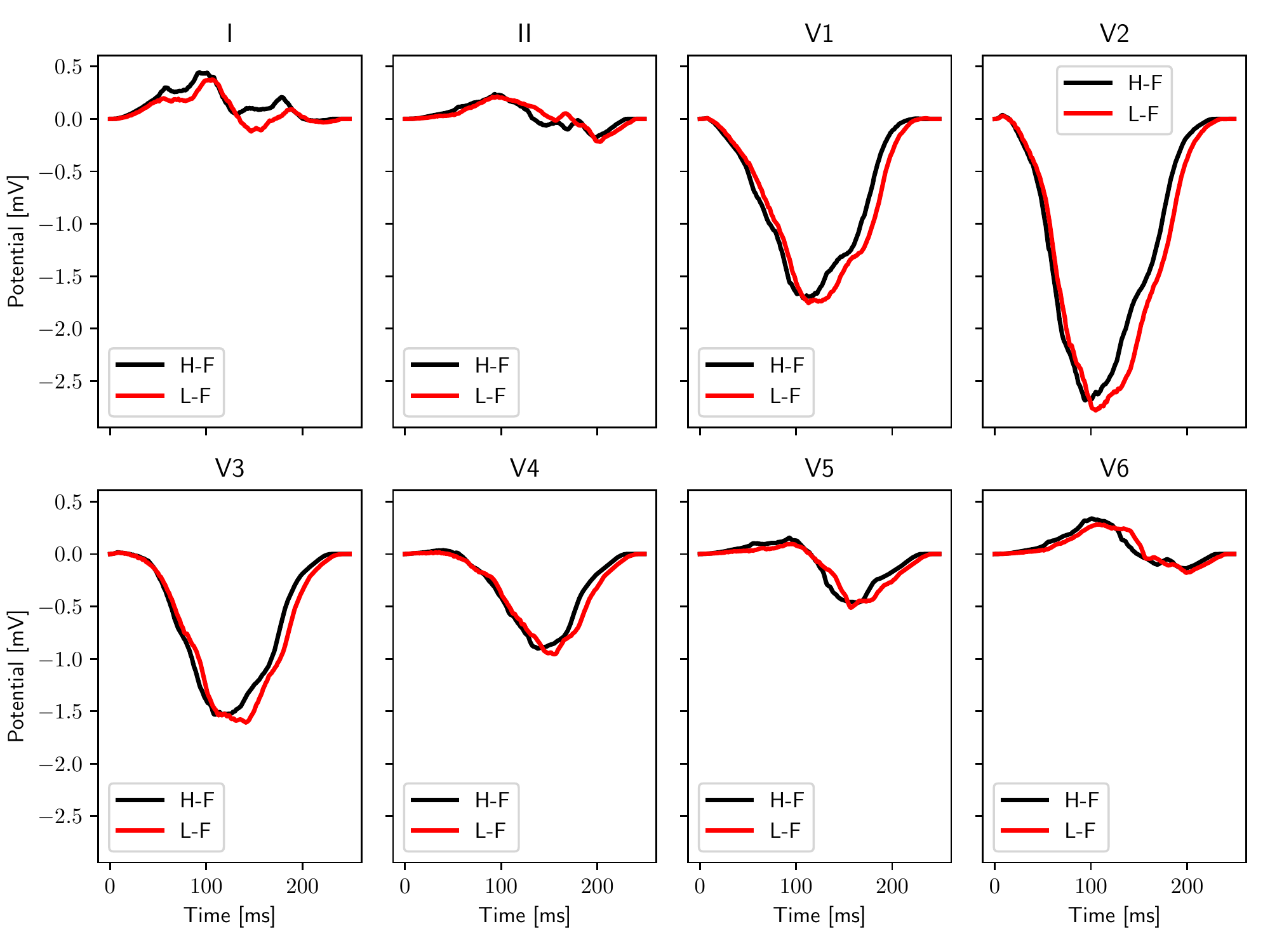}
\caption{Simulated ECGs for the HF and LF models (8 derivations). The ECGs correspond to a stimulation from the ground truth site (mid free wall).}
\label{fig:ecgcompare}
\end{figure}

In the MF case, Pearson's correlation between the HF and the LF ECG at
the ground truth was 0.98, as shown in Figure \ref{fig:ecgcompare}.
The computational cost for the LF model was roughly
9-fold lower than the HF model (runtime respectively equal to $0.6\,\mbox{s}$
and $5.6\,\mbox{s}$.)

\begin{figure}[htb]
\includegraphics[width=\columnwidth]{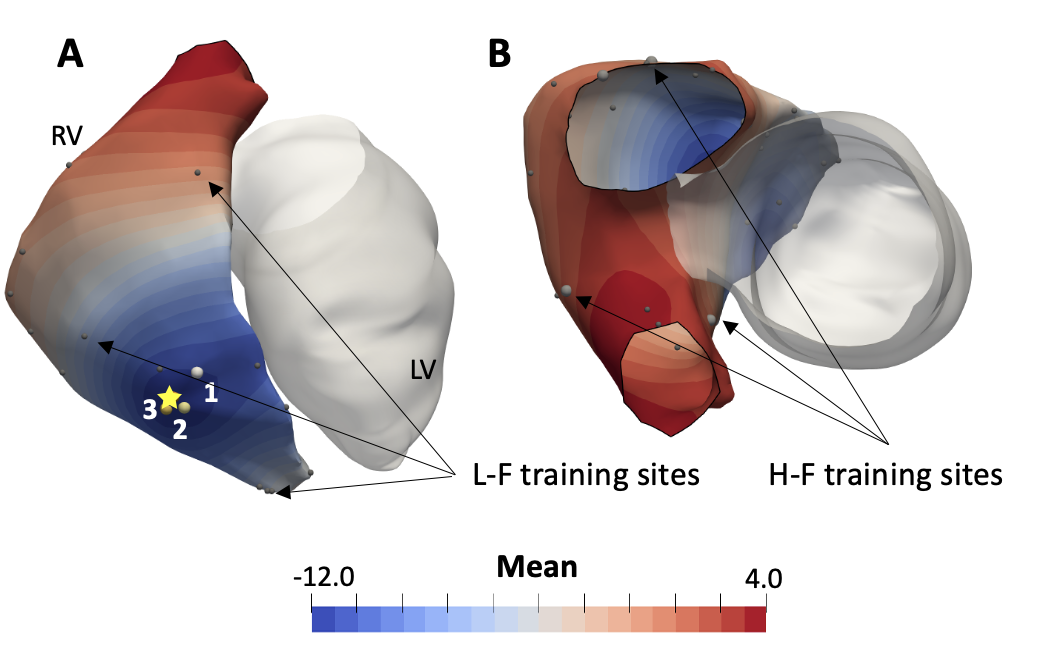}
\includegraphics[width=\columnwidth]{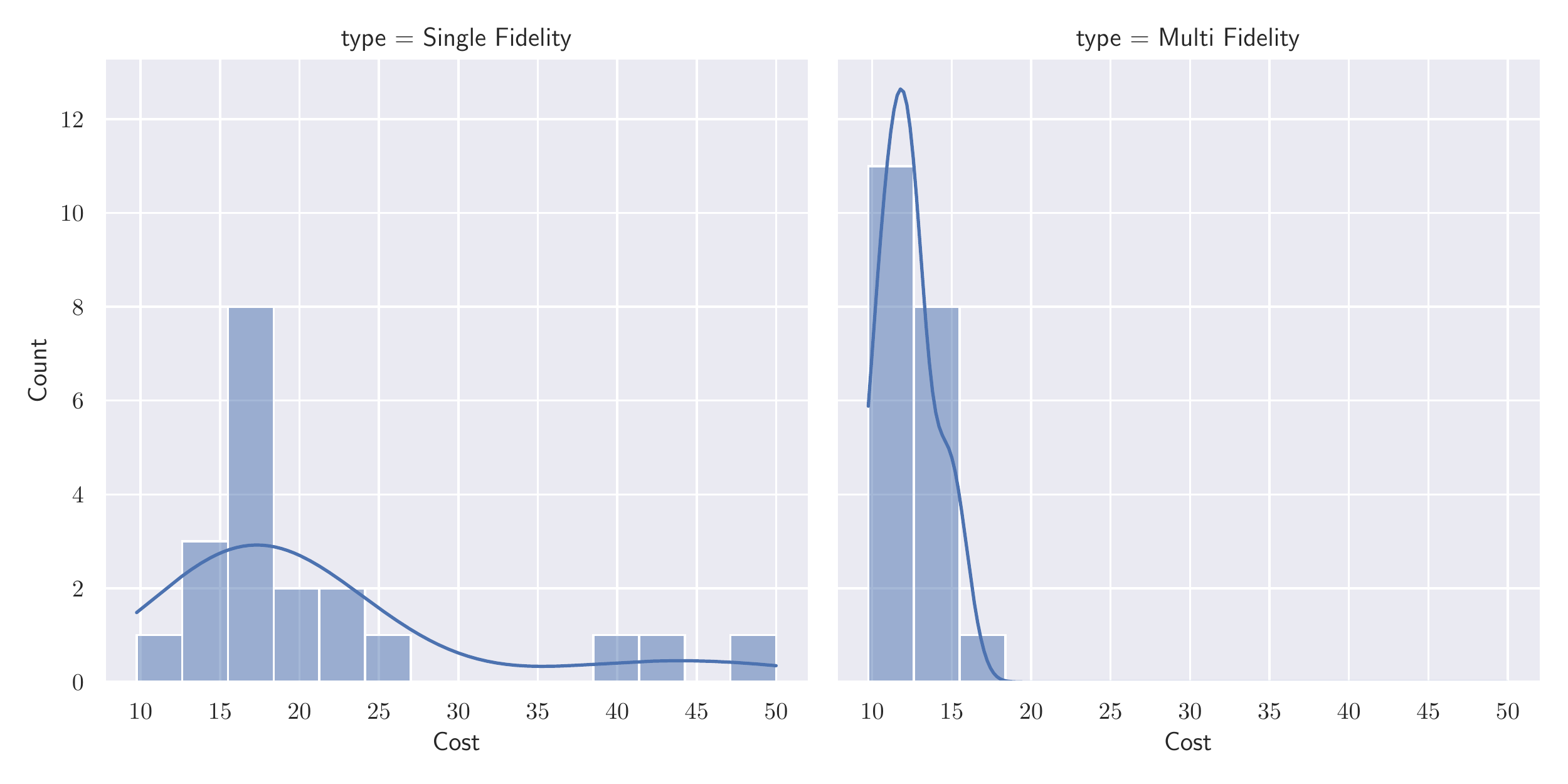}
\caption{Multi-fidelity experiment. Top row (A-B): two views of the mean of the GP, with iteration steps and training sites. Bottom row: randomized study on the training set ($N=20$), showing the distribution of the cost (one unit is an evaluation of the H-F model).}
\label{fig:multi}
\end{figure}

The MF framework converged in only 3 iterations,
thus highlighting the benefit of LF samples. A specific example is reported
in Figure~\ref{fig:multi}.
In order to asses the influence the training data, we performed
a experiment with randomized training set (20 runs in total).
The result is shown in Figure~\ref{fig:multi}, bottom panel.
The cost (one unit cost corresponds to a HF evaluation) of MF
is about half of SF (median respectively 17 and 11, inter-quartile
range 6 and 2), and we observed high cost runs (more than 20 iterations)
only the SF case.

\section{Discussion}
\label{sec:disc}

In this work, we have presented a novel methodology to identify an EAS from the ECG. We take advantage of state of the art machine learning tools---Gaussian process on manifolds---to build a surrogate model of the objective function and use Bayesian optimization to minimize it. In our results, we have shown that we need very few evaluations of our model to find the minimum in the SF case, and this is exacerbated in the MF case. In our tests, we able to solve the minimization problem in less than minute, which paves the way towards the translation of this tool to a clinical setting.  Intriguingly, the MF framework could be applied real-time with an high fidelity model corresponding to the real patient. For instance, our framework could be adapted to solve the problem of localizing ectopic foci during endocardial mapping, a tedious process usually performed by the clinician by manually pacing tentative locations.

We are also able to take advantage of inexpensive approximations of our model by building a MF GP surrogate model. We observed that this approach robustly reduced the median computational cost of 35\%, but more importantly the MF scheme reduced the variability in the cost by 3-fold.

One benefit of this method compared to existing ones, is the fact that
is we rely on global optimization, here possible because the dimension of the parameter space is small. In this way, the global minimum is readily available from the GP representing the loss function. The EAS localization problem is in fact potentially non-convex and hence prone to multiple local minima. The use a criterion that balance exploration and exploitation, such as the lower confidence bound, is key to find the global minimum in this problem more robustly.

The proof of concept presented in this work open some new research directions. So far, we only apply the optimization criterion to the high fidelity model in the MF setting. We could develop new rules that could acquire either the low or the high fidelity model, reducing the computational cost even further. We have also tested our methodology only with synthetic data, and we would need to verify its performance with patient data. Finally, we have only considered one EAS, but in practice there could be multiple, nearly simultaneous, activations at different locations. We plan to extend this framework to identify multiple sites from the same data.

In summary, we have presented a novel methodology to identify the EASs from non-invasive signals in cardiac electrophysiology. We hope this method will enable more precise interventions to treat cardiac arrhythmias.


\bibliography{ifacconf}                            

\end{document}